  \RequirePackage{fix-cm}
    \documentclass[twocolumn]{svjour3}          
  \smartqed  
  \usepackage{placeins}
  \usepackage{graphicx}
  \usepackage{appendix}
  \usepackage[english]{babel}
  \usepackage{verbatim}
  \usepackage{pifont}
  \newcommand{\cmark}{\ding{51}}%
  \newcommand{\xmark}{\ding{55}}%
  \usepackage[inline]{enumitem}
  \usepackage{graphicx}
  \usepackage{natbib}
  \usepackage{optidef}
  \usepackage{amsmath}
  \usepackage{amssymb}
  \usepackage{bm}
  \usepackage{algorithm, algorithmic}

  \usepackage{float}
  \usepackage{caption}
  \usepackage{subcaption}
  \usepackage{pgfplots}
  \usepackage{svg}
  \usepackage{multirow}
  \usepackage{tikzscale}
  \usetikzlibrary{calc}
  \usetikzlibrary{decorations,decorations.markings,decorations.text}
  \usetikzlibrary{patterns}
  \usetikzlibrary{arrows}
  \pgfdeclarepatternformonly{north east lines wide}{\pgfqpoint{-1pt}{-1pt}}{\pgfqpoint{10pt}{10pt}}{\pgfqpoint{9pt}{9pt}}%
  {
    \pgfsetlinewidth{0.4pt}
    \pgfpathmoveto{\pgfqpoint{0pt}{0pt}}
    \pgfpathlineto{\pgfqpoint{9.1pt}{9.1pt}}
    \pgfusepath{stroke}
  }
  \tikzset{%
      body/.style={inner sep=0pt,outer sep=0pt,shape=rectangle,draw,thick},
      marked body/.style={inner sep=0pt,outer sep=0pt,shape=rectangle,draw,pattern=north east lines},
      dimen/.style={<->,>=latex,thin,every rectangle node/.style={fill=white,midway,font=\sffamily}},
      symmetry/.style={dashed,thin},
  }
  
  \usepackage{makecell}
  \usepackage{hyperref}

\begin{document}

\title{Robust and stochastic compliance-based topology optimization with finitely many loading scenarios}


\titlerunning{Robust compliance optimization}        

\author{Mohamed Tarek         \and
        Tapabrata Ray
}

\institute{Mohamed Tarek \at
              UNSW Canberra, Northcott Drive, Campbell, ACT 2600 \\
              \email{m.mohamed@student.adfa.edu.au}           
           \and
           Tapabrata Ray \at
              UNSW Canberra, Northcott Drive, Campbell, ACT 2600 \\
              \email{t.ray@adfa.edu.au}
}

\date{Received: date / Accepted: date}

\maketitle

\begin{abstract}
  In this paper, the problem of load uncertainty in compliance problems is addressed where the uncertainty is described in the form of a set of finitely many loading scenarios. Computationally more efficient methods are proposed to exactly evaluate and differentiate: 1) the mean compliance, or 2) any scalar-valued function of the individual load compliances such as the weighted sum of the mean and standard deviation. The computational time complexities of all the proposed algorithms are analyzed, compared with the naive approaches and then experimentally verified. Finally, a mean compliance minimization problem, a risk-averse compliance minimization problem and a maximum compliance constrained problem are solved to showcase the efficacy of the proposed algorithms. The maximum compliance constrained problem is solved using the augmented Lagrangian method and the method proposed for handling scalar-valued functions of the load compliances, where the scalar-valued function is the augmented Lagrangian function.


\end{abstract}

\section{Introduction}

  \subsection{Optimization under data uncertainty}

    Every topology optimization problem has some input data, i.e. non-decision parameters, such as the load applied or material properties. The optimal solution depends on the value of the problem's data where a change in the data can cause a significant change in the objective value or render the optimal solution obtained infeasible. Robust optimization (RO), stochastic optimization (SO), risk-averse optimization (RAO) and reliability-based design optimization (RBDO) are some of the terms used in optimization literature to describe a plethora of techniques for handling uncertainty in the data of an optimization problem. 

    RO describes the problem's data using an uncertainty set \citep{Bertsimas2011}. The set can be continuous, discrete or a mixed set. The main characteristic of RO problems is that the constraints are required to be feasible for every data point in the uncertainty set. For more on RO, the readers are referred to \cite{Bertsimas2011} and \cite{AharonBen-Tal2009}.

    SO and RAO assume that the data follows a known probability distribution \citep{Shapiro2009,Choi2007}. Let $\bm{f}$ be a random load and $\bm{x}$ be the topology design variables. A probabilistic constraint can be defined as $P(g(\bm{x}; \bm{f}) \leq 0) \geq \eta$ where $\bm{f}$ follows a known probability distribution. This constraint is often called a chance constraint or a reliability constraint in RBDO. The objective of an SO problem is typically either deterministic or some probabilistic function such as the mean of a function of the random variable, its variance, standard deviation or a weighted sum of such terms.

    RAO can be considered a sub-field of SO which borrows concepts from risk analysis in mathematical economics to define various risk measures and tractable approximations to be used in objectives and/or constraints in SO. One such risk measure is the conditional value-at-risk (CVaR) \citep{Shapiro2009}. Other more traditional risk measures include the weighted sum of the mean and variance of a function or the weighted sum of the mean and standard deviation. For more on SO and RAO, the reader is referred to \cite{Shapiro2009}.

    RBDO and its ancestor, reliability analysis, are more commonly found in the sizing optimization literature. Classically, RBDO has been about solving optimization problems with a probabilistic constraint, called the reliability constraint, much like SO. One of the most common RBDO techniques used in topology optimization literature is the first-order reliability method (FORM). In FORM, the random variable $\bm{f}$ is assumed to be a function of a multivariate unit Gaussian random variable $\bm{u}$ relying on linearization and a Gaussian approximation of the probabilistic function's output. This approximation approach is known as the first-order second-moment (FOSM) approach. The choice of the linearization point $\bm{u}_0$ affects the accuracy of FOSM, where the mean $\bm{0}$ is typically outperformed by the less obvious alternative known as the most probable point (MPP) $\bm{u}^*$. There are two ways to define the MPP point: the reliability index approach (RIA) \citep{Chang1996,Tu1999} and the performance measure approach (PMA) \citep{Tu1999}. For more on RBDO and reliability analysis, the reader is referred to \cite{Choi2007} and \cite{Youn2004}. While classic RBDO has been about handling probabilistic reliability constraints, more recently the non-probabilistic RBDO (NRBDO) was developed, applying similar techniques as in classic RBDO but for handling set-based, non-probabilistic uncertainty to solve RO problems \citep{Luo2009,Kang2009,Guo2015,Zheng2018a,Wang2019a,Wang2019b}.

    In topology optimization literature, the term "\textit{robust topology optimization}" is often used to refer to minimizing the weighted sum of the mean, and variance or standard deviation of a function subject to probabilistic uncertainty \citep{Dunning2013,Zhao2014,Cuellar2018}. However, this use of the term "\textit{robust optimization}" is not consistent with the standard definition of RO in optimization theory literature, e.g. Ben-Tal et al. \cite{AharonBen-Tal2009}. The more compliant term is \textit{stochastic topology optimization} or \textit{risk-averse topology optimization}.

    The vast majority of works in literature on handling load uncertainty assume the load follows a probability distribution or lies in a continuous uncertainty set. In practice if a number of loading scenarios are known from sensor data, there is no way to use this data to perform risk-averse or robust compliance-based topology optimization. In this paper, computationally efficient approaches are proposed to compute and differentiate the mean compliance, its standard deviation and any scalar valued function of individual load compliances where each compliance is computed from a particular loading scenario. These approaches can then be used in risk-averse compliance minimization as well as handling robust compliance constraints where the uncertainty is described in the form of a set of finite loading scenarios.

  \subsection{Solid isotropic material with penalization}

    In this paper, the solid isotropic material with penalization (SIMP) method \citep{Bendsoe1989,Sigmund2001,Rojas-Labanda2015} is used to solve the topology optimization problems. Let $0 \leq x_e \leq 1$ be the decision variable associated with element $e$ in the ground mesh and $\bm{x}$ be the vector of such decision variables. Let $\rho_e$ be the pseudo-density of element $e$, and $\bm{\rho}(\bm{x})$ be the vector of such variables after sequentially applying to $\bm{x}$:
    \begin{enumerate}
      \item A chequerboard density filter typically of the form $f_1(\bm{x}) = \bm{A} \bm{x}$ for some constant matrix $\bm{A}$ \citep{Bendsoe2004}, 
      \item An interpolation of the form $f_2(y) = (1 - x_{min})y + x_{min}$ applied element-wise for some small $x_{min} > 0$ such as $0.001$, 
      \item A penalty such as the power penalty $f_3(z) = z^p$ applied element-wise for some penalty value $p$, and
      \item A projection method such as the regularized Heaviside projection \citep{Guest2004} applied element-wise.
    \end{enumerate}
    The compliance of the discretized design is defined as: $C = \bm{u}^T\bm{K}\bm{u} = \bm{f}^T\bm{K}^{-1}\bm{f}$ where $\bm{K}$ is the stiffness matrix, $\bm{f}$ is the load vector, and $\bm{u} = \bm{K}^{-1}\bm{f}$ is the displacement vector. The relationship between the global and element stiffness matrices is given by $\bm{K} = \sum\limits_e \rho_e \bm{K}_e$ where $\bm{K}_e$ is the hyper-sparse element stiffness matrix of element $e$ with the same size as $\bm{K}$.

  \subsection{Mean compliance minimization}

    A number of works in literature tackled the problem of load uncertainty in compliance minimization problems. Table \ref{tab:summary1} summarizes the literature on mean compliance minimization subject to different types of uncertainty. Nearly all the algorithms in literature assume the load follows a known probability distribution. Of all the works reviewed, only one work (\cite{Zhang2017a}) dealt with data-driven design with no distribution assumed. Zhang et. al. assumed the load can be any one of a finite number of loading scenarios where the loading scenarios can be collected from data. The algorithm proposed by Zhang et. al. can be trivially modified to handle weighted mean compliance which can be used in cases where the number of random variables are many following the approach by \cite{Zhao2014} or when the number of terms of the Karhunen-Loeve (K-L) expansion is high. However, the main limitation of this approach is that it can only be used to minimize the mean compliance which is not risk-averse since at the optimal solution, the compliance can still be very high for some probable load scenarios even if the mean compliance is minimized. This is one of the problems addressed in this work.

    \begin{table*}[h!]
     \centering
     \caption{Summary of literature on mean compliance minimization.}
     \begin{tabular}{| m{2cm} | m{4cm} | m{10cm}|} 
      \hline
      Paper & Uncertainty type & Summary \\
      \hline
      \cite{Guest2008} & Load components and load locations as distributions with arbitrary covariance & General scheme for handling uncertain loads and load location. Derived an approximately equivalent load distribution to result in the same compliance as the random node location. \\
      \hline
      \cite{Dunning2011} & Concentrated load magnitude and direction as distributions & Derived efficient formulations for the mean compliance, where the number of linear systems to be solved scales linearly with the number of independent random variables. Assumes that the distribution of the forces' magnitudes and rotations are known and independent. \\
      \hline
      \cite{Zhao2014} & Concentrated load magnitude and direction and distributed load as distributions or random fields & Requires fewer linear system solves per independent random variable compared to the approach by \cite{Dunning2011}. Demonstrated how the same approach can handle distributed load uncertainty, modeled as a stochastic field, using K-L expansion. \\ 
      \hline
      \cite{Zhang2017a} & Finite set of load scenarios & Proposed an efficient way to compute the mean compliance. Developed a randomized algorithm inspired by Hutchinson's trace estimator \cite{Hutchinson1990} to minimize the mean compliance leading to significant computational savings compared to the naive approach. \\
      \hline
      \cite{Liu2018a} & Fuzzy load & Modeled the load uncertainty using the fuzzy set theoretic cloud model. \\
      \hline
     \end{tabular}
     \label{tab:summary1}
    \end{table*}

  \subsection{Risk-averse compliance minimization}

    Some authors studied risk-averse compliance minimization by considering the weighted sum of the mean and variance, the weighted sum of the mean and standard deviation, as well as other risk measures. Table \ref{tab:summary2} summarizes the literature on risk-averse compliance minimization. All the works reviewed assumed the load to follow a known distribution or random field with a known covariance kernel. A number of works used the K-L expansion to handle the uncertainty when described using a random field. In all the works which use K-L expansion and sampling-based uncertainty propagation, the number of linear system solves can be made independent from the number of sampling points given the linearity assumption of the displacement as a function of the load exploited by \cite{Zhao2014b} in their derivation, even though in some of the works this property was not exploited. The number of linear system solves can therefore be assumed to be equal to the number of terms in the K-L expansion only, not the sampling points. There are no reports to the authors' knowledge on handle risk-averse compliance minimization when there is a finite set of loading scenarios, i.e. distribution-free. This work addresses this problem.

    \begin{table*}[h!]
     \centering
     \caption{Summary of literature on risk-averse compliance minimization.}
     \begin{tabular}{| m{2cm} | m{2cm} | m{12cm}|} 
      \hline
      Paper & Uncertainty type & Summary \\
      \hline
      \cite{Dunning2013} & Load magnitudes as independent distributions & Derived an efficient formulation for the variance of the compliance. This was used to minimize a weighted sum of the mean and variance of the compliance. \\
      \hline
      \cite{Zhao2014b} & Load as a random field & Used K-L expansion to quantify and describe the randomness using a few random variables, and used Monte Carlo simulation to calculate the fourth moment of those random variables which is required for the efficient computation of the standard deviation of the compliance. Minimized the weighted sum of the mean and standard deviation of the compliance. \\
      \hline
      \cite{Chen2010} & Load and material properties as random fields & Used K-L expansion to reduce the random field's dimensionality followed by Gaussian quadrature sampling to generate a representative set of scenarios to formulate the mean and variance of the compliance. Used the level-set method to minimize the weighted sum of the mean compliance and its variance. \\
      \hline
      \cite{Martinez-Frutos2016} & Load as a random field & Used K-L expansion to reduce the random field's dimensionality followed by sparse grid sampling to generate a representative set of scenarios to formulate the mean and variance of the compliance. Developed a multi-GPU density-based topology optimization framework for the large-scale minimization of the weighted sum of the mean compliance and its variance. \\
      \hline
      \cite{Cuellar2018} & Load and material properties as random fields & Used K-L expansion for uncertainty quantification and Gaussian quadrature for sampling, and combined them with the non-intrusive polynomial chaos expansion (PCE) method to provide more accurate estimators for the mean and standard deviation of the compliance and their gradients. \\
      \hline
      \cite{Martinez-Frutos2018} & Load and material properties as random fields & Used K-L expansion and the non-intrusive PCE with sparse grid sampling for the quantification and propagation of the uncertainty in the load and material properties. Minimized a different compliance risk measure called the \textit{excess probability}, which is the probability that the compliance exceeds a certain threshold value. \\
      \hline
      \cite{Garcia-Lopez2013} & Load as a distribution & Used multi-objective evolutionary optimization to optimize the mean and variance of the compliance and obtain the Pareto front of the two objectives. Used a sampling method for uncertainty propagation inspired from Taguchi's method for the design of experiments. In this case, the number of linear system solves is equal to the number of sampling points. That beside the use of an evolutionary algorithm which requires many evaluations of the mean and variance of the compliance make the computational cost of this approach extremely high even for medium-sized problems. \\
      \hline
      \cite{Kriegesmann2019} & Load as a distribution & Used FOSM instead of sampling to efficiently propagate the uncertainty estimating the mean and standard deviation of the compliance and their gradients from the means and standard deviations of the loads. A weighted sum of the mean and standard deviation of the compliance was then minimized. This approach assumes that the compliance is a linear function of the random load centered at the MPP load, an assumption which leads to a prediction error in the mean and standard deviation of the compliance. \\
      \hline
     \end{tabular}
     \label{tab:summary2}
    \end{table*}

  \subsection{Probabilistic constraints and reliability-based topology optimization}

    RBDO offers a number of techniques for efficient, approximate uncertainty propagation which can be used for handling probabilistic constraints involving compliance or otherwise. Most papers handling probabilistic constraints used methods from RBDO. Table \ref{tab:summary3} summarizes the literature on probabilistic constraint handling methods that can be used in compliance-based problems.

    \begin{table*}[h!]
     \centering
     \caption{Summary of literature on probabilistically constrained (or reliability constrained) compliance-based optimization.}
     \begin{tabular}{| m{2cm} | m{14cm}|} 
      \hline
      Paper & Summary \\
      \hline
      \cite{Keshavarzzadeh2017} & Solved the problem of volume minimization subject to a probabilistic compliance constraint. In one case, the authors constrained the mean compliance plus a multiple of its standard deviation which is equivalent to a reliability constraint assuming the compliance is normally distributed. In another, a reliability constraint was used such that the probability that the compliance exceeds a threshold value is constrained. Keshavarzzadeh et al. used the non-intrusive PCE and regularized Heaviside function to approximate the compliance reliability constraint and its gradient. PCE was also used to estimate the mean and standard deviation of the compliance and their gradients. \\
      \hline
      \cite{Kharmanda2002,Kharmanda2004} & Proposed the use of RBDO for topology optimization, also known as reliability-based topology optimization (RBTO), to handle probabilistic constraints due to random loads, geometry and material properties. \\
      \hline
      \cite{Jung2004} & Used FORM's PMA with SIMP to solve a volume minimization problem with a reliability constraint for geometrically nonlinear structures. \\
      \hline
      \cite{Kim2006} & Used FORM's RIA and PMA with SIMP to solve volume minimization problems with reliability constraints on the displacement and natural frequency of the structure under loading, material and geometry uncertainties. \\
      \hline
      \cite{Kim2007,Kim2008} & Used RIA and PMA together with evolutionary structural optimization (ESO) \citep{YM1992,XY1998,Huang2010a} to solve volume minimization problems with a reliability constraint subject to a random load and Young's modulus. \\
      \hline
      \cite{Ouyang2008} & Used FORM's RIA with the level-set method to solve a compliance minimization problem with a reliability constraint subject to uncertainty in the load and geometry of the ground mesh. \\
      \hline
      \cite{Silva2010} & proposed the use of an efficiently obtainable approximate MPP to avoid the need for solving the reliability or inverse reliability problems in every design iteration of RIA or PMA, respectively. \\
      \hline
      \cite{Silva2010, Nguyen2011} & Considered system reliability-based topology optimization, where an aggregated system failure probability is considered instead of component failure probabilities and component limit state functions. \\
      \hline
      \cite{Zhao2016} & Presented a comparison of a number of RBTO approaches to solve a few topology optimization problems including one with a compliance reliability constraint under stochastic load and Young's modulus. \\
      \hline
      \cite{Jalalpour2016} & Developed a bi-directional ESO (BESO) \citep{YM1992,XY1998,Huang2010a} algorithm for handling reliability constraints with displacement limit state functions and a finite number of probable loading scenarios in linearly elastic structures. \\
      \hline
      \cite{Yin2018} & Proposed an alternative RBTO approach using fuzzy set theory to describe the uncertainty. \\
      \hline
     \end{tabular}
     \label{tab:summary3}
    \end{table*}

  \subsection{Maximum compliance constraint}

    A number of works studied maximum compliance minimization and maximum compliance constrained problems under uncertain loading conditions. In these papers, the load was assumed to lie in a continuous uncertainty set, where no probability distribution is assumed. Therefore, they fall under the category of RO. A number of papers were also published on non-probabilistic reliability-based topology optimization (NRBTO) where new reliability indexes and performance measures are defined for various types of continuous uncertainty sets. While some of these works did not solve problems with maximum compliance constraints, the same techniques can be applied to handle maximum compliance constraints. Table \ref{tab:summary4} summarizes the literature on maximum compliance-constrained optimization including algorithms that can in theory be used to solve this class of problems. None of the reviewed papers handled the case of a finite number of loading scenarios instead of a continuous uncertainty set. This work addresses this issue.
 
    \begin{table*}[h!]
     \centering
     \caption{Summary of literature on maximum compliance constrained optimization.}
     \begin{tabular}{| m{1cm} | m{3cm} | m{11.5cm}|} 
      \hline
      Paper & Uncertainty type & Summary \\
      \hline
      \cite{Brittain2012} & Load vector with a fixed norm and arbitrary direction & Used a bi-level min-max optimization approach minimizing the objective with respect to the topology variables in the upper level problem, and maximizing with respect to the load in the lower level problem. However, an efficient algorithm was derived for the lower level maximization problem based on the KKT optimality conditions for the objective and the load's fixed-norm constraint. \\
      \hline
      \cite{Holmberg2015} & Load vector in a hyper-ellipsoid & Proposed a nonlinear semi-definite formulation to solve the set-maximum compliance minimization problem. \\
      \hline
      \cite{Thore2017} & Load vector in a hyper-ellipsoid & Generalized the approach from \cite{Holmberg2015} to handle maximum compliance and maximum stress constraints under the same assumption on the load vector. \\
      \hline
      \cite{Liu2018} & Multiple independent loads each in a hyper-ellipsoid & Proposed a bi-level formulation. Developed an efficient lower level algorithm by solving the Wolfe dual problem. The Wolfe dual problem of the lower level problem is a maximum generalized eigenvalue minimization problem which was solved using an iterative procedure. The multi-ellipsoidal uncertainty set generalizes the interval as well as the spherical uncertainty sets. \\
      \hline
      \cite{Luo2009} & Generic uncertain variables in a multi-ellipsoid set & Proposed an NRBDO reliability index and performance measure for handling non-probabilistic uncertainty. \\
      \hline
      \cite{Wang2018a} & Generic uncertain variables in an ellipsoid & Proposed another NRBDO reliability index for handling non-probabilistic uncertainty. \\
      \hline
      \cite{Wang2017,Wang2019a} & Generic uncertain variables in an interval & Proposed an NRBDO reliability index using interval arithmetic. \\
      \hline
      \cite{Zheng2018a} & Generic uncertain variables in multidimensional parallelepiped convex sets & Proposed an NRBDO reliability index and performance function. \\
      \hline
      \cite{Wang2019b} & Generic uncertain variables in a mixed interval and ellipsoidal set & Proposed an NRBDO reliability index. \\
      \hline
     \end{tabular}
     \label{tab:summary4}
    \end{table*}

  \subsection{Paper organization}

    The rest of this paper is organized as follows. The proposed approaches for handling load uncertainty in continuum compliance problems in the form of a large, finite number of loading scenarios are detailed in sections \ref{sec:proposed_mean}, \ref{sec:proposed_risk} and \ref{sec:proposed_max}. The experiments used and the implementations are then described in section \ref{sec:exp_impl}. Finally, the results are presented and discussed in section \ref{sec:results} before concluding in section \ref{sec:conclusion}.

\section{Compliance sample mean and its gradient} \label{sec:proposed_mean}

  \subsection{Naive approach}

    The compliance sample mean for a finite number $L$ of loading scenarios is $\mu_C = \frac{1}{L} \sum_{i=1}^L \bm{f}_i^T \bm{K}^{-1} \bm{f}_i$ where $\bm{f}_i$ is the $i^{th}$ load scenario, $\bm{K}$ is the stiffness matrix of the design and $\bm{F}$ is the matrix whose columns are the individual loading scenarios $\bm{f}_i$. The direct naive approach is to solve for $\bm{K}^{-1} \bm{f}_i$ for all $i$ and calculate the mean compliance using the above formula. This method is not efficient since it requires $L$ linear system solves plus some additional work to compute the mean with a time complexity of $O(L \times n_{dofs})$, where $n_{dofs}$ is the number of degrees of freedom in the design. When $\bm{F}$ is sparse with only a few $n_{loaded}$ degrees of freedom that are loaded, the complexity of the remaining work to compute the mean compliance $\frac{1}{L} \sum_{i=1}^L \bm{f}_i^T \bm{u}_i$ becomes $O(L \times n_{loaded})$. Even though the factorization of $\bm{K}^{-1}$ can be reused to solve for the $L$ linear systems, if $L$ is close to $n_{dofs}$, the complexity of solving for so many linear systems will be similar to that of the factorization, thus significantly adding to the running time. When using an iterative algorithm to solve for $\bm{K}^{-1}\bm{f}_i$, a good, but expensively formed, preconditioner such as the algebraic multi-grid preconditioner can be similarly reused. In general, significantly reducing the number of linear systems to solve is advantageous in practice even if, as theory may show, the running time is dominated by the initial linear system solve.

    Let the Jacobian of $\bm{\rho}(\bm{x})$ be $\nabla_{\bm{x}} \bm{\rho}(\bm{x})$. Let $\bm{u}_i$ be the displacement response due to load $\bm{f}_i$ and $C_i$ be the compliance $\bm{f}_i^T \bm{u}_i$. The stiffness matrix $\bm{K}$ is typically defined as: $\bm{K} = \sum_e \rho_e \bm{K}_e$. The partial derivative of the compliance $C_i$ with respect to $\rho_e$ is given by $\frac{\partial C_i}{\partial \rho_e} = -\bm{u}_i^T \bm{K}_e \bm{u}_i$. The gradient of $C_i$ with respect to the decision vector $\bm{x}$ is therefore given by: $\nabla_{\bm{x}} C_i(\bm{x}) = \nabla_{\bm{x}} \bm{\rho}(\bm{x})^T \nabla_{\bm{\rho}} C_i(\bm{\rho}(\bm{x}))$ where $\nabla_{\bm{\rho}} C_i(\bm{\rho}(\bm{x}))$ is the gradient of $C_i$ with respect to $\bm{\rho}$ at $\bm{\rho}(\bm{x})$. The gradient of the mean compliance $\mu_C$ is therefore given by $\nabla_{\bm{x}} \mu_C(\bm{x}) = \frac{1}{L} \sum_{i=1}^L \nabla_{\bm{x}} \bm{\rho}(\bm{x})^T \nabla_{\bm{\rho}} C_i(\bm{\rho}(\bm{x}))$. The additional complexity of computing the mean compliance and its gradient with respect to $\bm{\rho}$ is $O(n_E \times L)$. Note that the Jacobian of $\bm{\rho}(\bm{x})$ does not need to be formed explicitly to compute the gradient above, so long as there is a way to pre-multiply the Jacobian's transpose by a vector. The problem with the naive approach is it requires many linear system solves and so doesn't scale well to many loading scenarios.

  \subsection{Singular value decomposition}

    Less naively, one can first attempt to find the singular value decomposition (SVD) of $\bm{F}$. Let the compact SVD of the matrix $\bm{F}$ be $\bm{F} = \bm{U} \bm{S} \bm{V}^T$, where the number of non-zero singular values is $n_s$, $\bm{S}$ is the diagonal matrix of singular values, $\bm{U}$ is a $n_{dofs} \times n_s$ matrix with orthonormal columns, and $\bm{V}$ is $L \times n_s$ matrix with orthonormal columns. Given the SVD, the mean compliance can be written as: $\mu_C = \frac{1}{L} \sum_{i=1}^L \bm{f}_i^T \bm{K}^{-1} \bm{f}_i = \frac{1}{L} tr(\bm{F}^T \bm{K}^{-1} \bm{F})$. This can be further simplified:
    \begin{align}
     \frac{1}{L} tr(\bm{F}^T \bm{K}^{-1} \bm{F}) & = \frac{1}{L} tr(\bm{V} \bm{S} \bm{U}^T \bm{K}^{-1} \bm{U} \bm{S} \bm{V}^T) \\
     & = \frac{1}{L} tr(\bm{S} \bm{U}^T \bm{K}^{-1} \bm{U} \bm{S}) \\
     & = \frac{1}{L} \sum_{i=1}^{n_s} \bm{S}[i,i]^2 \times \bm{U}[:,i]^T \bm{K}^{-1} \bm{U}[:,i]
    \end{align}
    This method requires only $n_s$ linear system solves and an SVD. $n_s$ will be small if the loads in $\bm{F}$ are highly correlated or if only a few degrees of freedom are loaded, i.e. the loads are sparse. Let $n_{loaded}$ be the few loaded degrees of freedom. It is possible to prove in this case that the number of singular values $n_s \leq n_{loaded}$. The computational time complexity of computing the SVD of $\bm{F}$ in the dense case is $O(min(L, n_{dofs})^2 max(L, n_{dofs}))$, while in the sparse case it is only $O(n_{loaded}^2 L)$. If $n_{loaded}$ is a small constant, finding the SVD will be very efficient. Additionally, when only $n_{loaded}$ degrees of freedom are loaded in $\bm{F}$, only the same degrees of freedom will be non-zero in $\bm{U}$, therefore $\bm{U}$ will also be sparse. Other than the complexity of SVD, the additional work to compute the mean compliance has a computational time complexity of $O(n_s \times n_{dofs})$ when $\bm{F}$ (and $\bm{U}$) are dense, and $O(n_s \times n_{loaded})$ when $\bm{F}$ (and $\bm{U}$) are sparse.

    Given the efficient formula for the mean compliance and using the derivative rule of the inverse quadratic from the appendix, the partial $\frac{\partial \mu_C}{\partial \rho_e}$ is given by: \\ $-\frac{1}{L} \sum_{i=1}^{n_s} \bm{S}[i,i]^2 (\bm{K}^{-1}\bm{U})[:,i]^T \bm{K}_e (\bm{K}^{-1}\bm{U})[:,i]$. The time complexity of computing this assuming we already computed $\bm{K}^{-1} \bm{U}$ is $O(n_s \times n_E)$. 

  \begin{table*}[h!]
   \centering
   \caption{Summary of the computational cost of the algorithms discussed to calculate the mean compliance and its gradient. \#Lin is the number of linear system solves required.}
   \begin{tabular}{| m{1.7cm} | m{0.7cm} | m{0.8cm} | m{4cm} | m{4cm}|} 
    \hline
    \multirow{2}{3em}{Method} & \multirow{2}{2em}{\#Lin} & \multirow{2}{2em}{SVD?} & \multicolumn{2}{c|}{Time complexity of additional work} \\\cline{4-5}
    & & & Dense & Sparse \\
    \hline
    \hline
    Exact-Naive & \(L\) & \xmark & \(O(L \times (n_{dofs} + n_E))\) & \(O(L \times (n_{loaded} + n_E))\) \\
    \hline
    Exact-SVD & \(n_s\) & \cmark & \(O(n_s \times (n_{dofs} + n_E))\) & \(O(n_s \times (n_{loaded} + n_E))\) \\
    \hline
   \end{tabular}
   \label{tab:perf_mean}
  \end{table*}

\section{Scalar-valued function of load compliances and its gradient} \label{sec:proposed_risk}

  In this section, the above approach for computing the sample mean compliance will be generalized to handle the sample variance and standard deviations. The sample variance of the compliance $C$ is given by $\sigma_C^2 = \frac{1}{L-1} \sum_{i=1}^L (C_i - \mu_C)^2$. The sample standard deviation $\sigma_C$ is the square root of the variance. Let $\bm{C}$ be the vector of compliances $C_i$, one for each load scenario. In vector form, $\sigma_C^2 = \frac{1}{L-1} (\bm{C} - \mu_C \bm{1})^T (\bm{C} - \mu_C \bm{1})$. $\bm{C} = diag(\bm{A})$ is the diagonal of the matrix $\bm{A} = \bm{F}^T \bm{K}^{-1} \bm{F}$.

  \subsection{Naive approach}

    If one can compute the vector of load compliances $\bm{C}$, computing the variance and standard deviation is trivial. This approach requires $L$ linear system solves which can be computationally prohibitive if $L$ is large. Other than the linear system solves, the remaining work of computing $C_i = \bm{f}_i^T \bm{u}_i$ for all $i$ has a complexity of $O(L \times n_{dofs})$ when $\bm{F}$ is dense and $O(L \times n_{loaded})$ when $\bm{F}$ is sparse with only $n_{loaded}$ loaded degrees of freedom. In order to compute the vector $\bm{C}$ exactly, one needs to compute $\bm{u}_i = \bm{K}^{-1} \bm{f}_i$ for all $i$. These can further be used to compute the gradients of the load compliances $C_i$ which can be combined to form the Jacobian $\nabla_{\bm{\rho}} \bm{C}$. Assuming $\bm{u}_i$ is cached for all $i$, the time complexity of computing the Jacobian using $\frac{\partial C_i}{\partial \rho_e} = -\bm{u}_i^T \bm{K}_e \bm{u}_i$ is $O(n_E \times L)$.

    However, when interested in the gradient of a scalar-valued function $f$ of $\bm{C}$, there is no need to form the full Jacobian $\nabla_{\bm{x}} \bm{C}(\bm{x})$. It suffices to define an operator to pre-multiply an arbitrary vector $\bm{w}$ by $\nabla_{\bm{x}} \bm{C}(\bm{x})^T$. Using the chain rule, the gradient of $f$ with respect to $\bm{x}$ is given by $\nabla_{\bm{x}} f(\bm{C}(\bm{x})) = \nabla_{\bm{x}} \bm{C}(\bm{x})^T \nabla_{\bm{C}} f(\bm{C}(\bm{x}))$. This operator is equivalent to attempting to find the gradient of the weighted sum of $\bm{C}$, $\bm{w}^T \bm{C}$, where $\bm{w}$ is the constant vector of weights. In the case of a general scalar-valued function $f$, $\bm{w}$ would be $\nabla_{\bm{C}} f(\bm{C}(\bm{x}))$ and is treated as a constant. In the case of the variance, $f(\bm{C}) = \sigma_C^2 = \frac{1}{L-1} (\bm{C} - \mu_C \bm{1})^T (\bm{C} - \mu_C \bm{1})$, therefore $\bm{w} = \nabla_{\bm{C}} f(\bm{C}(\bm{x})) = \frac{2}{L-1} (\bm{C} - \mu_C \bm{1})$. And in the case of the standard deviation $\sigma_C$, $\bm{w} = \frac{1}{(L-1)\sigma_C} (\bm{C} - \mu_C \bm{1})$. This means that computing $\bm{C}$ is required to form $\bm{w}$.

    By caching $\bm{u}_i = \bm{K}^{-1}\bm{f}_i$ for all $i$ when computing $\bm{C}$, one can find the $e^{th}$ element of $\nabla_{\bm{x}} \bm{C}(\bm{x})^T \bm{w}$ using $\sum_{i=1}^L -w_i \bm{u}_i^T \bm{K}_e \bm{u}_i$, where $w_i$ is $i^{th}$ element of $\bm{w}$. Computing $\bm{u}_i^T \bm{K}_e \bm{u}_i$ requires constant time complexity, therefore the additional time complexity of computing $\nabla_{\bm{x}} \bm{C}(\bm{x})^T \bm{w}$ after computing $\bm{C}$ with the direct method is $O(L \times n_E)$. In this case, this is the same complexity as forming the Jacobian first and then multiplying, but in the next algorithms, it will be different.

  \subsection{Singular value decomposition}

    Much like in the mean compliance calculation, the SVD of $\bm{F}$ can be computed to find $C_i$ for all $i$ more efficiently from $\bm{K}^{-1} \bm{U} \bm{S}$. The number of linear system solves required to compute $\bm{K}^{-1} \bm{U} \bm{S}$ is $n_s$, the number of singular values of $\bm{F}$. The computational cost of computing $C_i = \bm{f}_i^T \bm{u}_i = \bm{f}_i^T (\bm{K}^{-1} \bm{U} \bm{S}) V^T[:,i]$ for all $i$ using $\bm{K}^{-1} \bm{U} \bm{S}$ and $\bm{V}$ is $O(L \times n_s \times n_{dofs})$ when $\bm{F}$ is dense and $O(L \times n_s \times n_{loaded})$ when $\bm{F}$ is sparse with only $n_{loaded}$ degrees of freedom loaded. The Jacobian $\nabla_{\bm{\rho}} \bm{C}$ can be built by first computing $\bm{K}^{-1} \bm{F}$ from the cached $\bm{K}^{-1} \bm{U} \bm{S}$ then using it much like in the exact method without SVD. This has a time complexity of $O((n_s \times n_{dofs} + n_E) \times L)$.

    However, when interested in $\nabla_{\bm{\rho}} \bm{C}^T \bm{w}$ instead, a more efficient approach can be used. Let $\bm{D}_{\bm{w}}$ be the diagonal matrix with the vector $\bm{w}$ on the diagonal.
    \begin{align}
     \nabla_{\bm{\rho}} \bm{C}^T \bm{w} & = \nabla_{\bm{\rho}} (\bm{C}^T \bm{w}) = \nabla_{\bm{\rho}} tr(\bm{D}_{\bm{w}} \bm{F}^T \bm{K}^{-1} \bm{F}) \\
     & = \nabla_{\bm{\rho}} tr(\bm{V}^T \bm{D}_{\bm{w}} \bm{V} \bm{S} \bm{U}^T \bm{K}^{-1} \bm{U} \bm{S})
    \end{align}
    Let $\bm{X} = \bm{V}^T \bm{D}_{\bm{w}} \bm{V}$ and $\bm{Q} = \bm{K}^{-1} \bm{U} \bm{S}$. The partial derivative of the above with respect to $\rho_e$ is:
    \begin{align}
     & \frac{\partial}{\partial \rho_e} tr(\bm{X} \bm{Q}^T \bm{S} \bm{U}^T \bm{K}^{-1} \bm{U} \bm{S}) = -tr(\bm{X} \bm{Q}^T \bm{K}_e \bm{Q})
    \end{align}

    Note that one can cache $\bm{Q} = \bm{K}^{-1} \bm{U} \bm{S}$ when finding the function value above to be reused in the sensitivity analysis. Let $\bm{Y}_e = \bm{Q}^T \bm{K}_e \bm{Q}$. The trace above is therefore $tr(\bm{X} \bm{Y}_e) = tr(\bm{X}^T \bm{Y}_e) = \sum_{i=1}^{n_s} \sum_{j=1}^{n_s} \bm{X}[i,j] \times \bm{Y}_e[i,j]$. Computing $\bm{Y}_e[i,j]$ from the pre-computed $\bm{Q}$ requires a constant time complexity for each element $e$, and computing $\bm{X}$ has a time complexity of $O(L \times n_s^2)$. The additional time complexity of computing $\nabla_{\bm{\rho}} \bm{C}^T \bm{w}$ using this method is therefore $O((n_E + L) \times n_s^2)$. So if $n_s \ll L$, significant computational savings can be made compared to directly computing the Jacobian then doing the matrix-vector multiplication $\nabla_{\bm{\rho}} \bm{C}^T \bm{w}$ which has a complexity of $O((n_s \times n_{dofs} + n_E) \times L)$.

    \begin{table*}[h!]
      \centering
      \caption{Summary of the computational cost of the algorithms discussed to calculate the load compliances $\bm{C}$ as well as $\nabla_{\bm{\rho}} \bm{C}^T \bm{w}$ for any vector $\bm{w}$. \#Lin is the number of linear system solves required. This can be used to compute the variance, standard deviation as well as other scalar-valued functions of $\bm{C}$. If the full Jacobian is required, the naive method requires the same computational cost as that of computing $\nabla_{\bm{\rho}} \bm{C}^T \bm{w}$, and the SVD-based method has a time complexity of $O((n_s \times n_{dofs} + n_E) \times L)$ for the additional work other than the linear system solves and SVD.}
      \begin{tabular}{|m{1.7cm} | m{0.7cm} | m{0.8cm} | m{5.0cm} | m{5.0cm} |} 
       \hline
       \multirow{2}{3em}{Method} & \multirow{2}{2em}{\#Lin} & \multirow{2}{2em}{SVD?} & \multicolumn{2}{c|}{Time complexity of additional work} \\\cline{4-5}
       & & & Dense & Sparse \\
       \hline
       \hline
       Exact-Naive & \(L\) & \xmark & \(O(L \times (n_{dofs} + n_E))\) & \(O(L \times (n_{loaded} + n_E))\) \\
       \hline
       Exact-SVD & \(n_s\) & \cmark & \(O(L \times n_s \times n_{dofs} + (n_E + L) \times n_s^2)\) & \(O(L \times n_s \times n_{loaded} + (n_E + L) \times n_s^2)\) \\
      \hline
     \end{tabular}
     \label{tab:perf_scalar}
    \end{table*}

\section{Maximum compliance constraint} \label{sec:proposed_max}

  The maximum compliance constraint can be efficiently handled using the augmented Lagrangian optimization algorithm \citep{Bertsekas1996}. Assume the following maximum compliance constrained problem is to be solved for some objective $g(\bm{x})$ using the augmented Lagrangian algorithm:
  \begin{mini!}|l|[3]
    {\bm{x}}{g(\bm{x})}{}{}
    \addConstraint {\quad C_i = \bm{f}_i^T \bm{K}^{-1} \bm{f}_i \leq C_t}{\quad \forall i \in 1 \dots L}
    \addConstraint {0 \leq x_e \leq 1}{\quad \forall e \in 1 \dots n_E}
  \end{mini!}
  where $C_t$ is the maximum compliance allowed. In the augmented Lagrangian algorithm, the problem is transformed as follows:
  \begin{mini!}|l|[3]
    {\bm{x}}{L(\bm{x}; \bm{\lambda}, r)}{}{}
    \addConstraint {0 \leq x_e \leq 1}{\quad \forall e \in 1 \dots n_E}
  \end{mini!}
  \begin{multline}
    L(\bm{x}; \bm{\lambda}, r) = g(\bm{x}) + \\
    \sum_{i=1}^{L} \biggl( \lambda_i (C_i - C_t) + r \max(C_i - C_t, 0)^2 \biggr)
  \end{multline}
  where $\bm{\lambda}$ is the vector of Lagrangian multipliers $\lambda_i$, one for each compliance constraint, and $r$ is the constant coefficient of the quadratic penalty. Solving the above problem using a first-order box constrained algorithm requires the gradient of $L(\bm{x})$. Writing $L(\bm{x})$ in vector form:
  \begin{align}
    L(\bm{x}) = g(\bm{x}) + \bm{\lambda}^T (\bm{C} - C_t \bm{1}) + r \bm{M}' \bm{M}
  \end{align}
  where $\bm{M}$ is the vector whose $i^{th}$ element is $max(C_i - C_t, 0)$. The gradient of $L(\bm{x})$ is given by:
  \begin{align}
    \nabla_{\bm{x}} L(\bm{x}) & = \nabla_{\bm{x}} g + \nabla_{\bm{x}} \bm{\rho}^T (\nabla_{\bm{\rho}} (\bm{\lambda}^T (\bm{C} - C_t \bm{1}) + r \bm{M}' \bm{M}) \\
    & = \nabla_{\bm{x}} g + \nabla_{\bm{x}} \bm{\rho}^T \nabla_{\bm{\rho}} \bm{C}^T (\bm{\lambda} + 2 \bm{M})
  \end{align}
  As shown in the previous sections, calculating the product $\nabla_{\bm{\rho}} \bm{C}^T (\bm{\lambda} + 2 \bm{M})$ can be done efficiently by finding the gradient $\nabla_{\bm{\rho}} (\bm{C}^T \bm{w})$ using $\bm{w} = (\bm{\lambda} + 2 \bm{M})$. Therefore, the results from Table \ref{tab:perf_scalar} apply.

\section{Setup and Implementation} \label{sec:exp_impl}

  In this section, the most important implementation details and algorithm settings used in the experiments are presented.

  \subsection{Test problems}

    \subsubsection{2D cantilever beam}
      \begin{figure}
        \centering
        \resizebox{0.5\textwidth}{!}{
          \begin{tikzpicture}
              \draw[fill,color=gray!70] (0,-0.1) rectangle (-0.3,2.1);
              \node [align=center, body,line width=1.2pt,minimum height=2cm,minimum width=8cm,anchor=south west] (body1) at (0,0) {};
              \draw (body1.south east) -- ++(1.6,0) coordinate (D1) -- +(5pt,0);
              \draw (body1.north east) -- ++(1.6,0) coordinate (D2) -- +(5pt,0);
              \draw [dimen] (D1) -- (D2) node {40mm};

              \draw (body1.north west) -- ++(0,2) coordinate (D1) -- +(0,5pt);
              \draw (body1.north east) -- ++(0,2) coordinate (D2) -- +(0,5pt);
              \draw [dimen] (D1) -- (D2) node {160mm};

              \draw[->,line width=1pt] (8,1) -- (9,1) -- (9,0.2);
              \node (arrowhead) at (8.7,0.25) {$\bm{F}_1$};

              \draw (body1.north west) -- ++(0,1.5) coordinate (D1) -- +(0,5pt);
              \draw (4,2) -- ++(0,1.5) coordinate (D2) -- +(0,5pt);
              \draw [dimen] (D1) -- (D2) node {80mm};

              \draw[->,line width=1pt] (3.4,2.6) -- (4,2);
              \node (arrowhead) at (3.2,2.8) {$\bm{F}_2$};

              \draw (body1.south west) -- ++(0,-1.5) coordinate (D1) -- +(0,5pt);
              \draw (6,0) -- ++(0,-1.5) coordinate (D2) -- +(0,5pt);
              \draw [dimen] (D1) -- (D2) node {120mm};

              \draw[->,line width=1pt] (6,0) -- (5.4,-0.6);
              \node (arrowhead) at (5.2,-0.8) {$\bm{F}_3$};

              \draw[->,line width=1pt] (0.3,2.3) -- (0.1,2);
              \draw[->,line width=1pt] (0.6,2.3) -- (0.8,2);
              \draw[->,line width=1pt] (1.2,2.3) -- (1.1,2);
              \draw[->,line width=1pt] (1.5,2.3) -- (1.8,2);
              \draw[->,line width=1pt] (2.2,2.3) -- (2.0,2);
              \draw[->,line width=1pt] (2.5,2.3) -- (2.8,2);
              \draw[->,line width=1pt] (3.1,2.3) -- (3.0,2);
              \draw[->,line width=1pt] (3.3,2.3) -- (3.5,2);
              \draw[->,line width=1pt] (4.5,2.3) -- (4.4,2);
              \draw[->,line width=1pt] (4.8,2.3) -- (4.9,2);
              \draw[->,line width=1pt] (5.3,2.3) -- (5.1,2);
              \draw[->,line width=1pt] (5.6,2.3) -- (5.9,2);
              \draw[->,line width=1pt] (6.2,2.3) -- (6.0,2);
              \draw[->,line width=1pt] (6.5,2.3) -- (6.9,2);
              \draw[->,line width=1pt] (7.2,2.3) -- (7.2,2);
              \draw[->,line width=1pt] (7.5,2.3) -- (7.7,2);
              \draw[->,line width=1pt] (8.1,2.3) -- (7.9,2);

              \draw[->,line width=1pt] (0.3,-0.3) -- (0.3,0);
              \draw[->,line width=1pt] (0.8,-0.3) -- (0.6,0);
              \draw[->,line width=1pt] (1.2,-0.3) -- (1.1,0);
              \draw[->,line width=1pt] (1.5,-0.3) -- (1.5,0);
              \draw[->,line width=1pt] (2.0,-0.3) -- (2.2,0);
              \draw[->,line width=1pt] (2.5,-0.3) -- (2.8,0);
              \draw[->,line width=1pt] (3.1,-0.3) -- (3.0,0);
              \draw[->,line width=1pt] (3.5,-0.3) -- (3.2,0);
              \draw[->,line width=1pt] (3.8,-0.3) -- (3.7,0);
              \draw[->,line width=1pt] (4.1,-0.3) -- (4.1,0);
              \draw[->,line width=1pt] (4.4,-0.3) -- (4.4,0);
              \draw[->,line width=1pt] (5.0,-0.3) -- (4.9,0);
              \draw[->,line width=1pt] (5.3,-0.3) -- (5.3,0);
              \draw[->,line width=1pt] (5.5,-0.3) -- (5.6,0);
              \draw[->,line width=1pt] (6.0,-0.3) -- (6.2,0);
              \draw[->,line width=1pt] (6.6,-0.3) -- (6.7,0);
              \draw[->,line width=1pt] (7.3,-0.3) -- (7.2,0);
              \draw[->,line width=1pt] (7.7,-0.3) -- (7.5,0);
              \draw[->,line width=1pt] (8.1,-0.3) -- (7.9,0);

              \draw[->,line width=1pt] (8.3,0.3) -- (8.0,0.1);
              \draw[->,line width=1pt] (8.3,0.5) -- (8.0,0.8);
              \draw[->,line width=1pt] (8.3,1.0) -- (8.0,1.2);
              \draw[->,line width=1pt] (8.3,1.5) -- (8.0,1.5);
              \draw[->,line width=1pt] (8.3,1.7) -- (8.0,1.9);
          \end{tikzpicture} \newline
        }
        \caption{Cantilever beam problem. $\bm{F}_2$ and $\bm{F}_3$ are at 45 degree angles.}
        \label{fig:CantBeam}
      \end{figure}
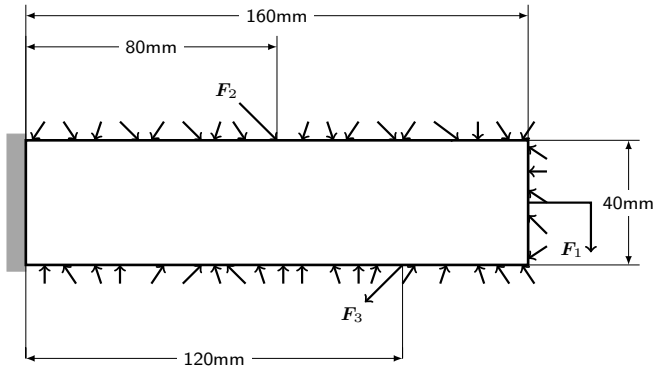

      The 2D cantilever beam problem shown in Figure \ref{fig:CantBeam} was used to run the experiments. A ground mesh of plane stress quadrilateral elements was used, where each element is a square of side length $1 \text{ mm}$, and a sheet thickness of $1 \text{ mm}$. Linear iso-parametric interpolation functions were used for the field and geometric basis functions. A Young's modulus of 1 MPa and Poisson's ratio of 0.3 were used. Finally, a chequerboard density filter for unstructured meshes was used with a radius of 2 mm \cite{Huang2010a}. A 3D version of the problem above was also solved. Details of the 3D problem and the results are shown in the appendix.

      Three variants of the cantilever beam problem were solved:
      \begin{enumerate}
        \item Minimization of the mean compliance $\mu_C$ subject to a volume constraint with a volume fraction of 0.4,
        \item Minimization of a weighted sum of the mean and standard deviation (mean-std) of the compliance $\mu_C + 2.0 \sigma_C$ subject to a volume constraint with a volume fraction of 0.4, and
        \item Volume minimization subject to a maximum compliance constraint with a compliance threshold of $70000 \text{ Nmm}$.
      \end{enumerate}
      A total of 1000 load scenarios were sampled from:
      \begin{align}
        \bm{f}_i = s_1 \bm{F}_1 + s_2 \bm{F}_2 + s_3 \bm{F}_3 + \frac{1}{7} \sum_{j=4}^{10} s_j \bm{F}_j
      \end{align}
      where $\bm{F}_1$, $\bm{F}_2$ and $\bm{F}_3$ are unit vectors with directions as shown in Figure \ref{fig:CantBeam}. $\bm{F}_2$ and $\bm{F}_3$ are at 45 degrees. $s_1$, $s_2$ and $s_3$ are identically and independently uniformly distributed random variables between -2 and 2. $\bm{F}_j$ for $j$ in $4 \dots 10$ are vectors with non-zeros at all the surface degrees of freedom without a Dirichlet boundary condition. The non-zero values are identically and independently normally distributed random variables with mean 0 and standard deviation 1. $s_j$ for $j$ in $4 \dots 10$ are also identically and independently normally distributed random variables with mean 0 and standard deviation 1. The same loading scenarios were used for the 3 test problems. Let $\bm{F}$ be the matrix whose columns are the sampled $\bm{f}_i$ vectors. The way the loading scenarios are defined, the rank of $\bm{F}$ can be at most 10 and was actually exactly 10 in our experiments. Given the low rank structure of $\bm{F}$, the SVD approaches should be expected to be significantly more efficient than their naive counterparts.

    \subsubsection{3D cantilever beam}

      A 3D version of the 2D cantilever beam test problem above was also solved using the methods proposed. A 60 mm x 20 mm x 20 mm 3D cantilever beam was used with hexahedral elements of cubic shape and side length of 1 mm. The loads $\bm{F}_1$, $\bm{F}_2$ and $\bm{F}_3$ were positioned at (60, 10, 10), (30, 20, 10) and (40, 0, 10) where the coordinates represent the length, height and depth respectively. The remaining loads and multipliers were sampled from the same distributions as the 2D problem. A density filter radius of 3 mm was also used for the 3D problem.

  \subsection{Software}

    All the topology optimization algorithms described in this paper were implemented in TopOpt.jl \footnote{https://github.com/mohamed82008/TopOpt.jl} using the Julia programming language \citep{Bezanson2014} version 1.3 for handling generic unstructured, iso-parametric meshes.

  \subsection{Settings}

    The value of $x_{min}$ used was $0.001$ for all problems and algorithms. Penalization was done prior to interpolation to calculate $\bm{\rho}$ from $\bm{x}$. A power penalty function and a regularized Heaviside projection were used. All of the problems were solved using 2 continuation SIMP routines. The first incremented the penalty value from $p = 1$ to $p = 6$ in increments of 0.5. Then the Heaviside projection parameter $\beta$ was incremented from $\beta = 0$ to $\beta = 20$ in increments of 4 keeping the penalty value fixed at 6. An exponentially decreasing tolerance from $1e-3$ to $1e-4$ was used for both continuations.

    The mean and mean-std compliance minimization SIMP subproblems problems were solved using the method of moving asymptotes (MMA) algorithm \cite{Svanberg1987}. MMA parameters of $s_{init} = 0.5$, $s_{incr} = 1.1$ and $s_{decr} = 0.7$ were used as defined in the MMA paper with a maximum of 1000 iterations for each subproblem. The dual problem of the convex approximation was solved using a log-barrier box-constrained nonlinear optimization solver, where the barrier problem was solved using the nonlinear CG algorithm for unconstrained nonlinear optimization \citep{Nocedal2006} as implemented in Optim.jl \footnote{https://github.com/JuliaNLSolvers/Optim.jl} \citep{KMogensen2018}. The nonlinear CG itself used the line search algorithm from \cite{Hager2006} as implemented in LineSearches.jl \footnote{https://github.com/JuliaNLSolvers/LineSearches.jl}. The stopping criteria used was the one adopted by the KKT solver, IPOPT \citep{Wachter2006}. This stopping criteria is less scale sensitive than the KKT residual as it scales down the residual by a value proportional to the mean absolute value of the Lagrangian multipliers.

    The maximum compliance constrained SIMP subproblems were solved using a primal-dual augmented Lagrangian method \citep{Bertsekas1996}. The inequality constraints were relaxed resulting in a box constrained max-min primal-dual problem. A projected gradient descent algorithm was used for the primal and dual problems with a backtracking line search. The maximum step size of the line search was initialized to 1 and adapted to be 1.5 the step size of the previous line search for both the primal and dual problems. A total of 10 dual iterations were used with a maximum of 50 primal iterations per dual iteration. The IPOPT termination criteria above was also used here. To regularize the scale of the problem, the compliance constraints were divided by the maximum compliance at the full ground mesh design. A trust region of 0.1 was used. An initial quadratic penalty coefficient of 0.1 was used with a growth factor of 3 in every dual iteration. Finally, an initial solution of 1.0 for all the primal variables and 1 for all the Lagrangian multipliers was used.

  \subsection{Replication of Results}
    
    The instructions and codes needed to replicate the results in this paper are given in \url{https://github.com/mohamed82008/RobustComplianceCode}.

\section{Results and Discussion} \label{sec:results}

  \subsection{Speed comparison}

    Tables \ref{tab:time_mean} and \ref{tab:time_std} show the values computed for the mean compliance $\mu_C$ and its standard deviation $\sigma_C$ respectively together with the time required to compute their values and gradients using: the exact naive approach (Exact-Naive) and the exact method with SVD (Exact-SVD). As expected, the proposed exact SVD approach computes the exact mean compliance or its standard deviation and their gradient in a small fraction of the time it takes to compute them using the naive approaches.

    \begin{table}[h!]
     \centering
     \caption{The table shows the function values of $\mu_C$ computed using the naive exact method (Exact-Naive) and the exact method with SVD (Exact-SVD). The table also shows the time required to compute $\mu_C$ and its gradient in each case.}
     \begin{tabular}{|c|c|c|}
      \hline
      Method & $\mu_C$ (Nmm) & Time (s) \\
      \hline
      \hline
      Exact-Naive & 3328.7 & 24.2 \\
      \hline
      Exact-SVD & 3328.7 & 0.4 \\
      \hline
     \end{tabular}
     \label{tab:time_mean}
    \end{table}

    \begin{table}[h!]
     \centering
     \caption{The table shows the function values of $\sigma_C$ and its gradients for a full ground mesh computed using the naive exact method (Exact-Naive) and the exact method with SVD (Exact-SVD). The table also shows the time required to compute $\sigma_C$ and its gradient in each case.}
     \begin{tabular}{|c|c|c|}
      \hline
      Method & $\sigma_C$ (Nmm) & Time (s) \\
      \hline
      \hline
      Exact-Naive & 4172.8 & 28.0 \\
      \hline
      Exact-SVD & 4172.8 & 1.5 \\
      \hline
     \end{tabular}
     \label{tab:time_std}
    \end{table}

  \subsection{Optimization}

\begin{figure}
      \centering
      \includegraphics[width=0.5\textwidth]{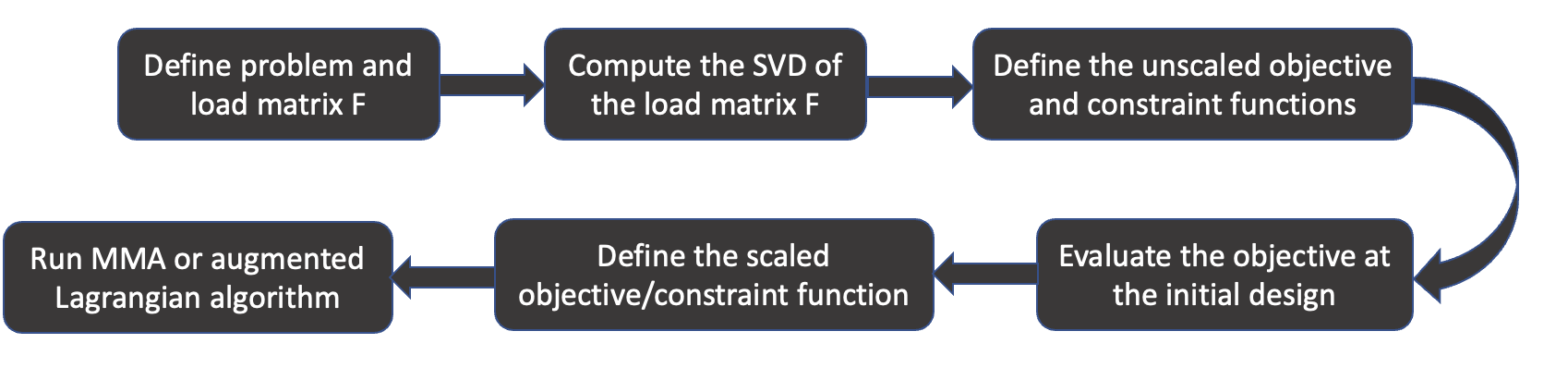}
      \caption{Flowchart of the experiments' workflow. Only the mean compliance objective, mean-std compliance objective or maximum compliance constraint are scaled by the inverse of their initial value. The volume function is not scaled.}
      \label{fig:flowchart}
    \end{figure}

    In this section, a number of stochastic, risk-averse and robust compliance-based optimization problems are solved using the proposed methods. Figure \ref{fig:flowchart} shows the experiments' workflow.

    \subsubsection{Mean compliance minimization}

\begin{figure}
        \centering
        \includegraphics[width=0.5\textwidth]{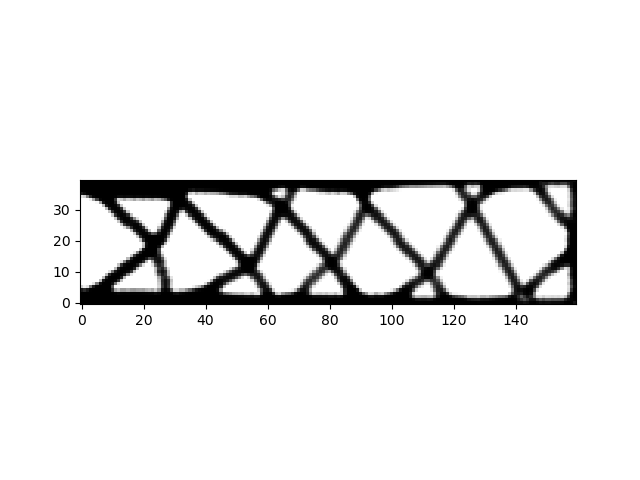}
        \caption{Optimal topology of the mean compliance minimization problem using continuation SIMP and the SVD-based method for evaluating the mean compliance.}
        \label{fig:mean}
      \end{figure}

\begin{figure}
        \begin{subfigure}[t]{0.45\textwidth}
          \centering
          \includegraphics[width=1\textwidth]{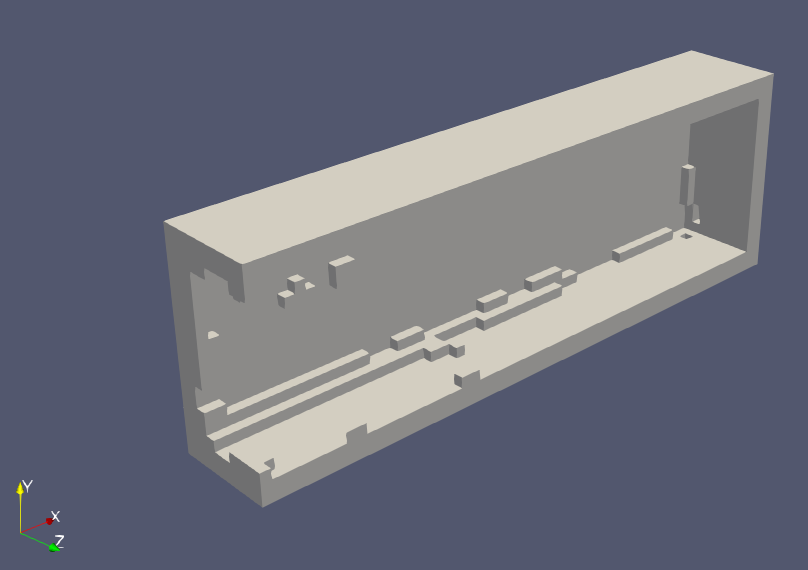}
          \caption{Left half}
        \end{subfigure}
        \begin{subfigure}[t]{0.45\textwidth}
          \centering
          \includegraphics[width=1\textwidth]{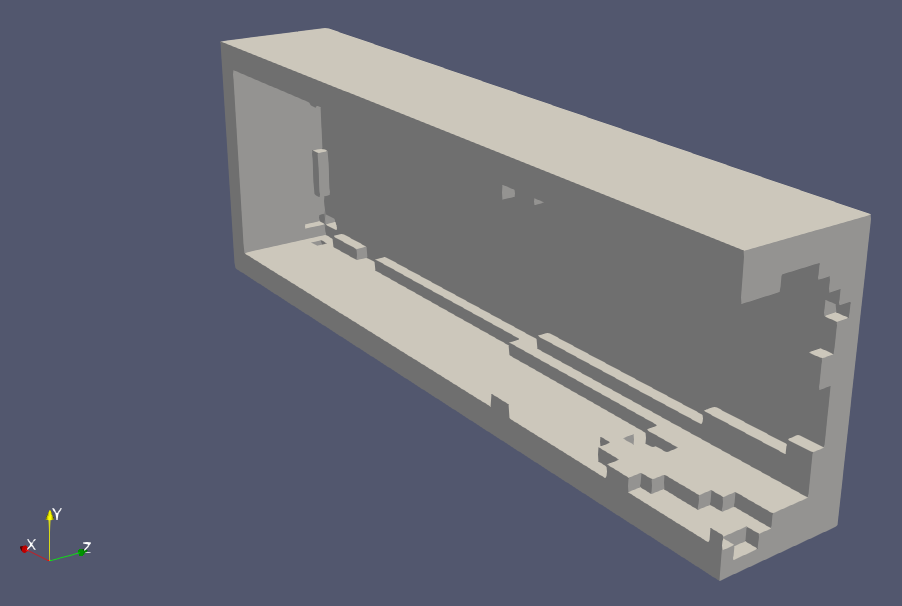}
          \caption{Right half}
        \end{subfigure}
        \caption{Cut views of the optimal topologies of the 3D mean compliance minimization problem using exact method with SVD.}
        \label{fig:exact_mean_3d}
      \end{figure}

      To demonstrate the effectiveness of the proposed approaches, the 2D and 3D cantilever beam problems described in section \ref{sec:exp_impl} were solved using the proposed SVD-based methods. Table \ref{tab:mean_stats} shows the statistics of the final optimal solutions obtained by minimizing the mean compliance subject to the volume fraction constraint using the SVD-based method to evaluate the mean compliance. The optimal topologies are shown in Figures \ref{fig:mean} and \ref{fig:exact_mean_3d}.

      \begin{table}[h!]
       \centering
       \caption{Summary statistics of the load compliances of the optimal solutions of the 2D and 3D mean compliance minimization problems using the SVD-based method to evaluate the mean compliance.}
       \begin{tabular}{|c|c|c|}
        \hline
        Compliance Stat & 2D & 3D \\
        \hline
        \hline
        $\mu_C$ (Nmm) & 9392.8 & 22072.1 \\
        \hline
        $\sigma_C$ (Nmm) & 9688.8 & 16628.7 \\
        \hline
        $C_{max}$ (Nmm) & 125455.0 & 184055.0 \\
        \hline
        $C_{min}$ (Nmm) & 467.9 & 1785.8 \\
        \hline
        $V$ & 0.400 & 0.400 \\
        \hline
        $Time$ (s) & 491.5 & 3849.6 \\
        \hline
       \end{tabular}
       \label{tab:mean_stats}
      \end{table}

    \subsubsection{Mean-std compliance minimization}

\begin{figure}
        \centering
        \includegraphics[width=0.5\textwidth]{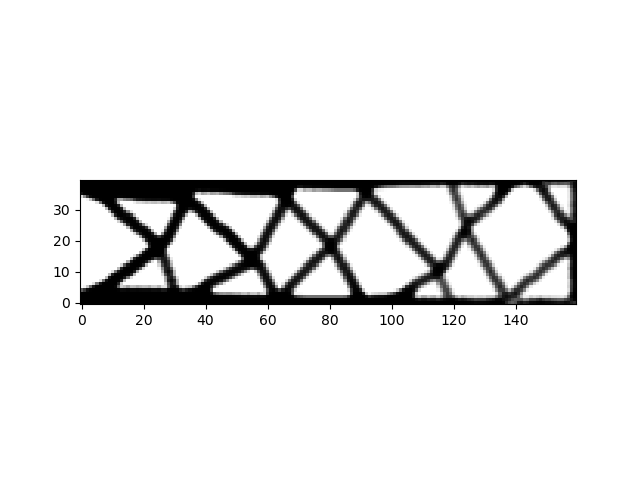}
        \caption{Optimal topology of the mean-std compliance minimization problem using continuation SIMP and the SVD-based method to compute the mean-std.}
        \label{fig:mean_std}
      \end{figure}

\begin{figure}
        \begin{subfigure}[t]{0.45\textwidth}
          \centering
          \includegraphics[width=1\textwidth]{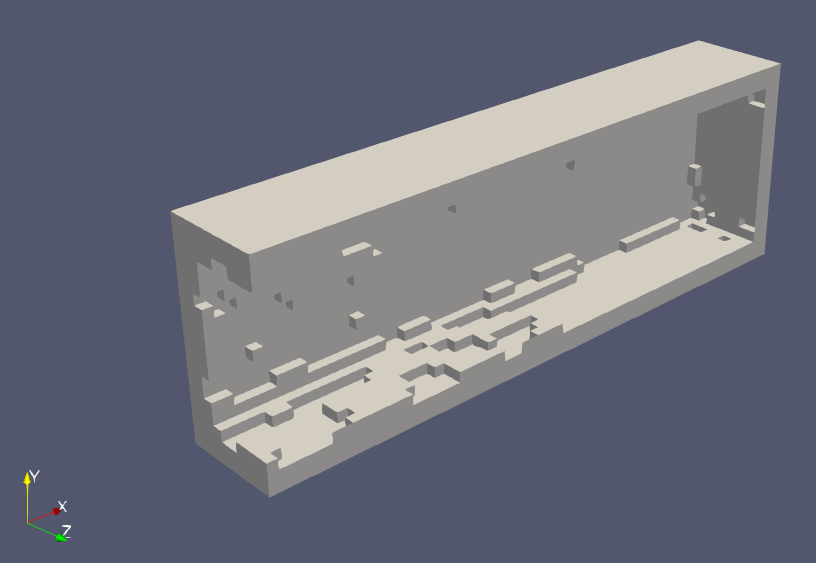}
          \caption{Left half}
        \end{subfigure}
        \begin{subfigure}[t]{0.45\textwidth}
          \centering
          \includegraphics[width=1\textwidth]{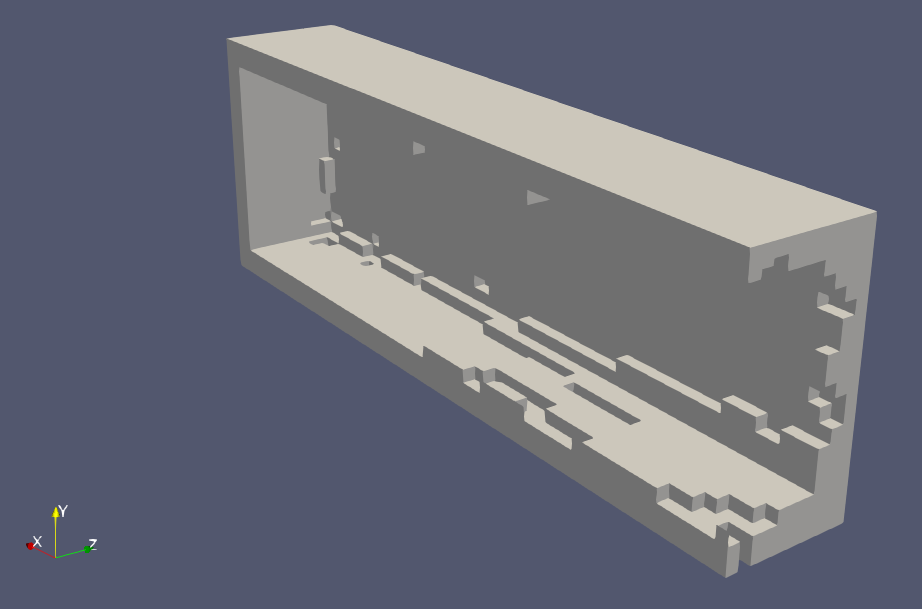}
          \caption{Right half}
        \end{subfigure}
        \caption{Cut views of the optimal topologies of the 3D mean-std compliance minimization problem using the exact method with SVD.}
        \label{fig:exact_mean_std_3d}
      \end{figure}

      Similarly, Table \ref{tab:mean_std_stats} shows the statistics of the final solutions of the 2D and 3D mean-std minimization problems solved using the SVD-based method. The optimal topologies are shown in Figures \ref{fig:mean_std} and \ref{fig:exact_mean_std_3d}. The algorithm converged to reasonable, feasible designs. Additionally, as expected the mean-std minimization algorithm converged to solutions with lower compliance standard deviations but higher mean compliances compared to the mean minimization algorithm.

      \begin{table}[h!]
       \centering
       \caption{Summary statistics of the load compliances of the optimal solutions of the 2D and 3D mean-std compliance minimization problems using the SVD-based method to evaluate the mean-std compliance.}
       \begin{tabular}{|c|c|c|}
        \hline
        Compliance Stat & 2D & 3D \\
        \hline
        \hline
        $\mu_C (Nmm) $ & 9796.9 & 22216.7 \\
        \hline
        $\sigma_C (Nmm) $ & 9240.0 & 16220.2 \\
        \hline
        $\mu_C + 2.0 \sigma_C (Nmm) $ & 28283.7 & 54848.8 \\
        \hline
        $C_{max}$ (Nmm) & 117883.1 & 176153.2 \\
        \hline
        $C_{min}$ (Nmm) & 527.7 & 1872.0 \\
        \hline
        $V$ & 0.400 & 0.400 \\
        \hline
        Time (s) & 229.8 & 3528.2 \\
        \hline
       \end{tabular}
       \label{tab:mean_std_stats}
      \end{table}

\begin{figure}
        \centering
        \includegraphics[width=0.5\textwidth]{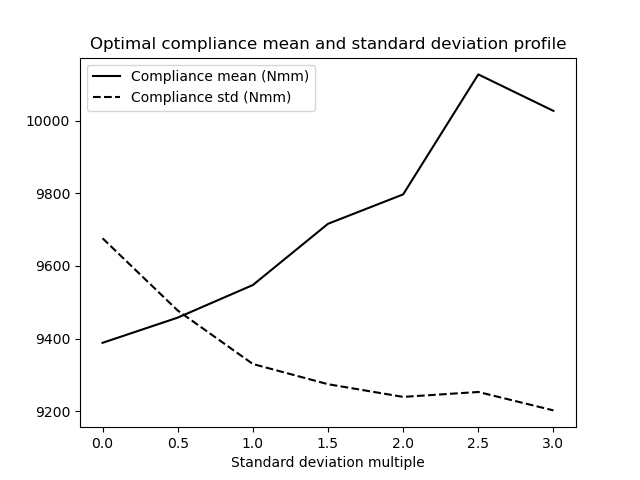}
        \caption{Profile of the optimal mean and standard deviation of the compliance for different standard deviation multiples in the objective.}
        \label{fig:mean_std_profile}
      \end{figure}

\begin{figure*}
        \begin{subfigure}[t]{0.3\textwidth}
          \centering
          \includegraphics[width=1\textwidth]{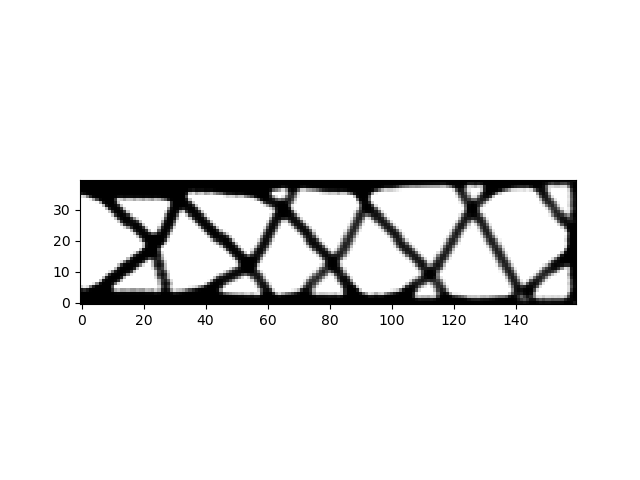}
          \caption{$m = 0$}
        \end{subfigure}
        \begin{subfigure}[t]{0.3\textwidth}
          \centering
          \includegraphics[width=1\textwidth]{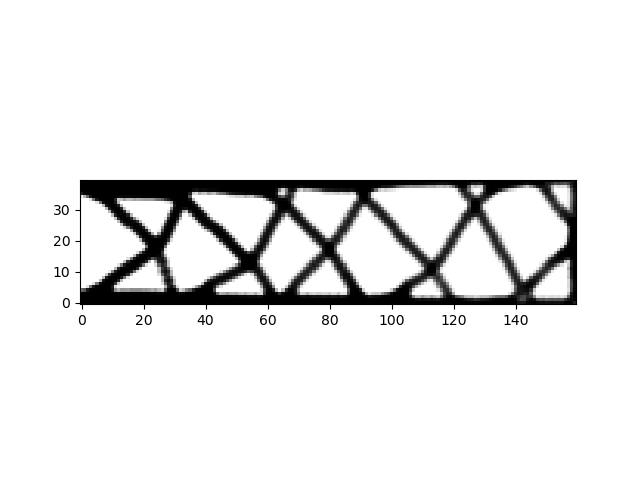}
          \caption{$m = 0.5$}
        \end{subfigure}
        \begin{subfigure}[t]{0.3\textwidth}
          \centering
          \includegraphics[width=1\textwidth]{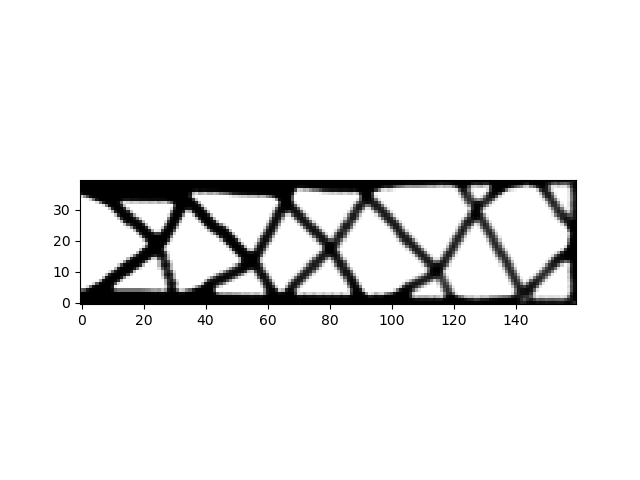}
          \caption{$m = 1.0$}
        \end{subfigure}
        \begin{subfigure}[t]{0.3\textwidth}
          \centering
          \includegraphics[width=1\textwidth]{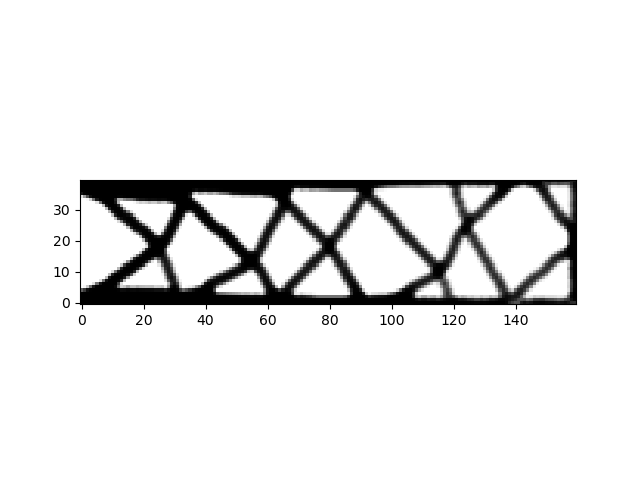}
          \caption{$m = 1.5$}
        \end{subfigure}
        \begin{subfigure}[t]{0.3\textwidth}
          \centering
          \includegraphics[width=1\textwidth]{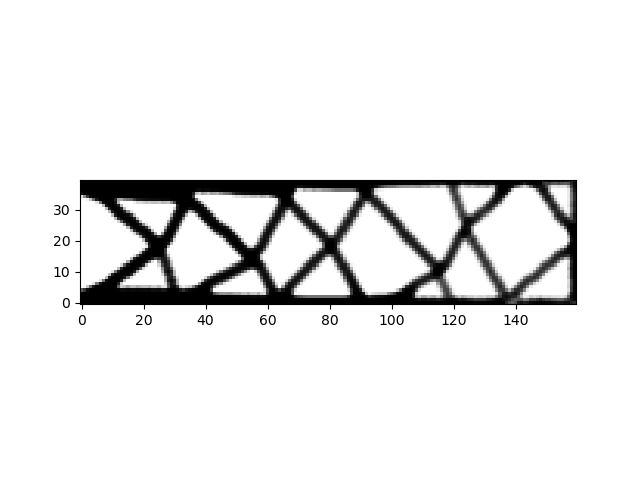}
          \caption{$m = 2.0$}
        \end{subfigure}
        \begin{subfigure}[t]{0.3\textwidth}
          \centering
          \includegraphics[width=1\textwidth]{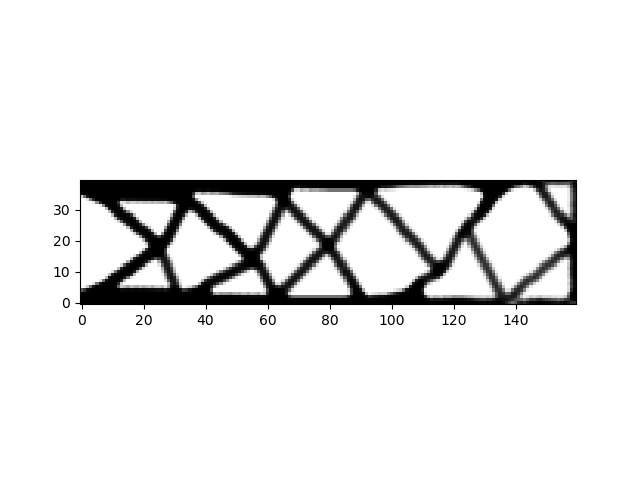}
          \caption{$m = 2.5$}
        \end{subfigure}
        \begin{subfigure}[t]{0.3\textwidth}
          \centering
          \includegraphics[width=1\textwidth]{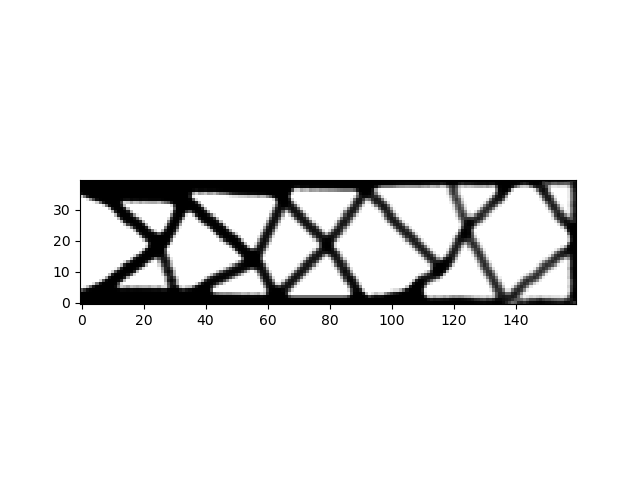}
          \caption{$m = 3.0$}
        \end{subfigure}
        \caption{Optimal topologies of the 2D mean-std compliance minimization problem using different standard deviation multiples $m$ in the objective $\mu_C + m \sigma_C$.}
        \label{fig:mean_m_std_2d}
      \end{figure*}

      To highlight the effect of the multiple $m$ of the standard deviation in the objective $\mu_C + m \times \sigma_C$, the same problem was solved for different values of $m$. Figure \ref{fig:mean_std_profile} shows the profile of the mean and standard deviation of the compliance. Interestingly due to the non-convexity of the problem, increasing the standard deviation's multiple can sometimes lead to a simultaneous increase or reduction in the mean and standard deviation of the compliance. The different optimal topologies are shown in Figure \ref{fig:mean_m_std_2d}.

    \subsubsection{Maximum compliance constrained optimization}

\begin{figure}
        \centering
        \includegraphics[width=0.5\textwidth]{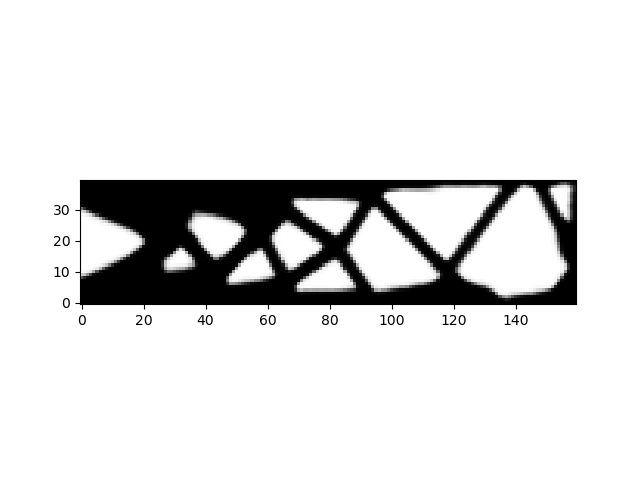}
        \caption{Optimal topology of the volume minimization problem subject to a maximum compliance constraint using continuation SIMP and the augmented Lagrangian method with the exact SVD approach. The maximum compliance of the design above is 69847.0 Nmm and the volume fraction is 0.584.}
        \label{fig:max}
      \end{figure}

\begin{figure}
        \centering
        \begin{subfigure}[t]{0.45\textwidth}
          \centering
          \includegraphics[width=1\textwidth]{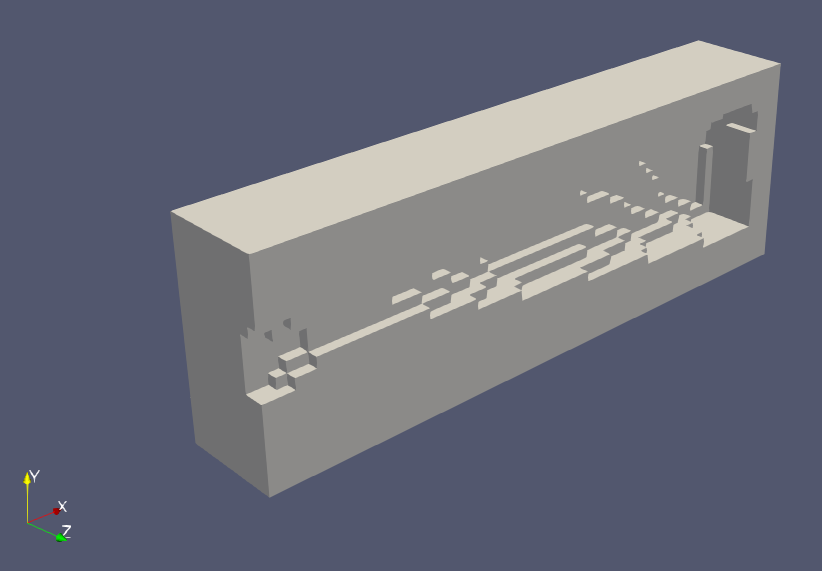}
          \caption{Left half}
        \end{subfigure}
        \begin{subfigure}[t]{0.45\textwidth}
          \centering
          \includegraphics[width=1\textwidth]{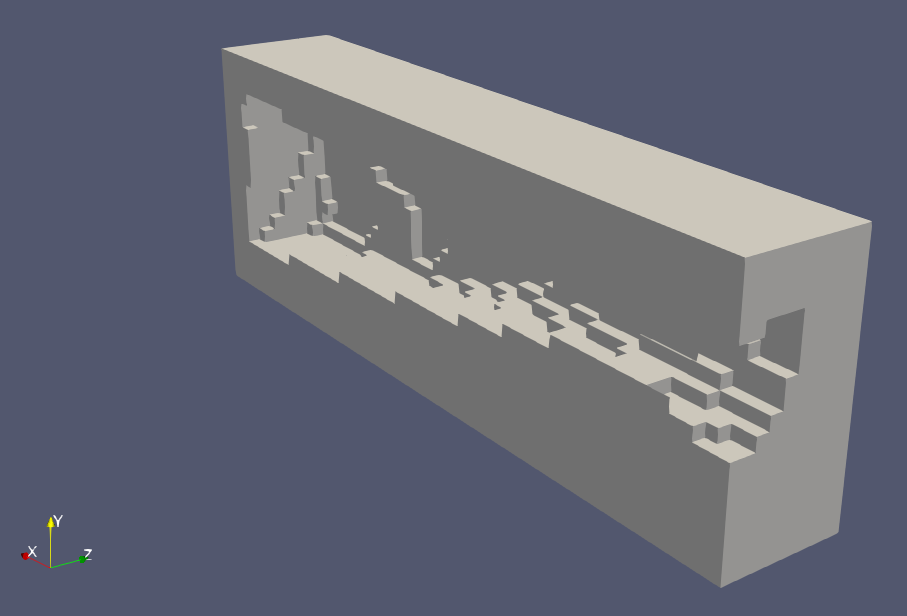}
          \caption{Right half}
        \end{subfigure}
        \caption{Cut views of the 3D optimal topology of the volume minimization problem subject to a maximum compliance constraint using continuation SIMP and the augmented Lagrangian method with the exact SVD approach. The maximum compliance of the design above is 68992.4 Nmm and the volume fraction is 0.791.}
        \label{fig:max_3d}
      \end{figure}

      The 2D and 3D maximum compliance constrained volume minimization problems were solved using the SVD-based approach. The 2D optimal topology, shown in Figure \ref{fig:max}, had a volume fraction of 0.584 and a maximum compliance of 69847.0 Nmm and was reached in 662.7 s. The 3D optimal topology, shown in Figure \ref{fig:max_3d}, had a volume fraction of 0.791 and a maximum compliance of 68992.4 Nmm and was reached in 43740.6 s.

\section{Conclusion} \label{sec:conclusion}

  In this paper, a number of exact methods were proposed to handle load uncertainty in compliance topology optimization problems where the uncertainty is described in the form of a set of finitely many loading scenarios. By exploiting low rank structures in loading scenarios, significant performance improvements were achieved using novel SVD-based methods. Such improvement was demonstrated via complexity analysis and computational experiments. The methods presented here are fundamentally data-driven in the sense that no probability distributions or domains are assumed for the loading scenarios. This sets this work apart from most of the literature in the domain of stochastic and robust topology optimization where a distribution or domain is assumed. Additionally, the methods proposed here were shown to be particularly suitable with the augmented Lagrangian algorithm when dealing with maximum compliance constraints.

\section{Acknowledgments}

  This research did not receive any specific grant from funding agencies in the public, commercial, or not-for-profit sectors.

\section{Conflict of Interest}

  The authors have no conflict of interest to declare.

\newpage

\appendix
\section{Partial derivative of the inverse quadratic form}

In this section, it will be shown that the $i^{th}$ partial derivative of:
\begin{align}
  f(\bm{x}) & = \bm{v}^T (\bm{A}(\bm{x}))^{-1} \bm{v}
\end{align}
is
\begin{align}
  \frac{\partial f}{\partial x_i} & = -\bm{y}^T \frac{\partial \bm{A}}{\partial x_i} \bm{y}^T
\end{align}
where $\bm{A}$ is a matrix-valued function of $\bm{x}$, $\bm{v}$ is a constant vector and $\bm{y} = \bm{A}^{-1} \bm{v}$ is a an implicit function of $\bm{x}$ because $\bm{A}$ is a function of $\bm{x}$.

\begin{align}
  \bm{v} & = \bm{A} \bm{y} \\
  \bm{0} & = \bm{A} \frac{\partial \bm{y}}{\partial x_i} + \frac{\partial \bm{A}}{\partial x_i} \bm{y} \\
  \frac{\partial y}{\partial x_i} & = - \bm{A}^{-1} \frac{\partial \bm{A}}{\partial x_i} \bm{y} \\
  f(\bm{x}) & = \bm{v}^T \bm{A}^{-1} \bm{v} \\
  & = \bm{y}^T \bm{A} \bm{y} \\
  \frac{\partial f}{\partial x_i} & = 2 \bm{y}^T \bm{A} \frac{\partial \bm{y}}{\partial x_i} + \bm{y}^T \frac{\partial \bm{A}}{\partial x_i} \bm{y} \\
  & = - 2 \bm{y}^T \bm{A} \bm{A}^{-1} \frac{\partial \bm{A}}{\partial x_i} \bm{y} + \bm{y}^T \frac{\partial \bm{A}}{\partial x_i} \bm{y} \\
  & = - 2 \bm{y}^T \frac{\partial \bm{A}}{\partial x_i} \bm{y} + \bm{y}^T \frac{\partial \bm{A}}{\partial x_i} \bm{y} \\
  & = - \bm{y}^T \frac{\partial \bm{A}}{\partial x_i} \bm{y}
\end{align}

\newpage


%
%

\bibliographystyle{spbasic}      
\bibliography{references}   

%
%

\end{document}